\documentclass[12pt]{iopart}
\usepackage{iopams}  
\usepackage{graphicx}
\begin{document}

\title{Weak magnetism phenomena in heavy-fermion superconductors: selected $\mu$SR studies
}

\author{A.~Amato\ddag\footnote[3]{To
whom correspondence should be addressed (alex.amato@psi.ch)}, M.J.~Graf\dag, A.~de~Visser$\|$, H.~Amitsuka\P, D.~Andreica$^+$ and A.~Schenck$\sharp$
}

\address{\ddag\ Laboratory for Muon-Spin Spectroscopy, Paul Scherrer Institute, 
CH-5232 Villigen PSI, Switzerland\\
\dag\ Department of Physics, Boston College, Chestnut Hill, MA 02467, USA\\
$\|$\ Van der Waals--Zeeman Institute, University of Amsterdam, NL-1018 XE Amsterdam, The Netherlands\\
\P\ Division of Physics, Graduate School of Science, Hokkaido University, Sapporo 060-0810, Japan\\
$^+$\ Faculty of Physics, Babes-Bolyai University, RO-3400 Cluj-Napoca, Romania\\
$\sharp$ Institute for Particle Physics, ETH Zurich, CH-5232 Villigen PSI, Switzerland
}

\begin{abstract}
The behavior of the so-called weak moment antiferromagnetic states, observed in the heavy-fermion superconductors UPt$_3$ and URu$_2$Si$_2$, is discussed in view of recent $\mu$SR results obtained as function of control parameters like chemical substitution and external pressure. 
In UPt$_3$, the Pd substitution for Pt reveals the dynamical character of the weak moment order. On the other hand, $\mu$SR measurements performed on samples in which Th substitutes U suggest that crystallographic disorder on the magnetic sites deeply affects the fluctuation timescale.
In URu$_2$Si$_2$, a phase separation between the so-called hidden order state, present at ambient pressure, and an antiferromagnetic state, occurring under pressure, is observed. In view of the pressure-temperature phase diagram obtained by $\mu$SR, it is deduced that the respective order parameters have different symmetries.
\end{abstract}


\submitto{\JPCM}

\maketitle

\section{Introduction}

The free motion of the conduction electrons is often considered to explain most of the physical properties in usual metals. In some systems, however, the interactions between electrons cannot be totally neglected. As a result, these systems are better described within Landau's Fermi 
liquid theory \cite{landau}, where there exists a one-to-one correspondence between the states of the interacting system and those of the free Fermi gas. The Fermi liquid is composed by quasiparticles with renormalized mass $m^*$.

A striking example of Fermi-liquid behavior is provided by $f$-electron systems called heavy fermions, where the interactions are so strong that the value of the effective mass of the quasiparticles is several orders of magnitude higher than the bare electron mass (see for example \cite{grewe} and references therein). The large $m^*$ value of these systems stems from the strong correlation between the localized $f$ moments and the conduction electrons. While at high temperature the $f$ electrons and conduction electrons interact weakly, at low temperature these two subsets of electrons become strongly coupled. Thus, at sufficiently low temperature, heavy-fermion compounds behave like systems of heavy itinerant electrons.

A fascinating aspect of this class of compounds is the observation that, within the heavy-fermion regime, a wealth of ground states can occur. Probably the most intriguing observation is the occurrence of a superconducting state, reported for several heavy-fermion systems and first detected in CeCu$_2$Si$_2$ by Steglich \etal \cite{steglich}. Two exciting aspects of this superconducting state are the facts that (i) it occurs in a system possessing local magnetic moments ({\it i.e.} the $f$-electrons) that are known to rapidly suppress superconductivity in conventional systems; and (ii) it appears as a co-operative phenomenon involving heavy quasiparticles that form Cooper pairs below $T_c$. This latter deduction is based for example on the observed scaling of the huge linear coefficient of the electronic specific heat ($\gamma = C_p /T$) with the specific-heat jump at $T_c$. 

Even at an early stage in the development of heavy-fermion physics, 
numerous experimental evidence was put forward which strongly suggested that superconductivity in these systems might be unconventional. Unconventional superconductivity could result from the nature of the mechanism providing the attractive force necessary for the Cooper-pair formation. In conventional superconductors, the electrons are paired in a spin-singlet zero-angular-momentum state ($L=0$), which results from the fact that their binding is described in terms of the emission and absorption of waves of lattice density. This isotropic state leads to the formation of a superconducting gap in the electronic excitations over the whole Fermi surface. On the other hand, 
heavy-fermion superconductivity is observed to show a close interplay with magnetic fluctuations. This seems to indicate that the attractive effective interaction between the electrons in the superconducting heavy-fermion systems is not provided by the electron-lattice interaction as in ordinary superconductors, but rather is mediated by electronic spin fluctuations. Recently, important measurements performed under pressure have provided a strong case for this picture (see for example \cite{mathur,saxena,huxley,pfleiderer,aoki}). In a number of systems, it was demonstrated that superconductivity occurs under pressure on the brink of magnetic order, that is when strong spin fluctuations are present throughout the sample. This non-conventional ({\it i.e.} non-BCS) mechanism is believed to lead to an unconventional configuration of the heavy-fermion superconducting state, which may involve anisotropic, nonzero-angular-momentum states ($L \ne 0$ {\it i.e.} $p$ or $d$-states; see \cite{sigrist} for a review and references therein). Hence, this unconventional superconducting state might be characterized by the presence on the Fermi surface of points or lines with vanishing superconducting gap, leading to power-law temperature dependencies below $T_c$ of excitation-dependent physical properties. Finally, one might stress that this unconventional superconducting state presents a close analogy to the combined hard-core repulsion spin-fluctuation-induced superfluidity of $^3$He, where the $p$-wave superfluidity is mediated by ferromagnetic spin fluctuations \cite{leggett}.

An additional puzzle is the observation in a number of systems of an apparent coexistence of static magnetism and heavy-fermion superconductivity. Microscopic studies, in particular making use of the capability of the $\mu$SR technique, demonstrated in some cases, as CeCu$_2$Si$_2$, the competition between both ground states, {\it i.e.} magnetism and superconductivity do not microscopically coexist and appear as two different, mutually exclusive ground states of the same subset of electrons \cite{luke, feyerherm}. Another class of heavy-fermion compounds exhibits a physical picture of two rather independent electron subsets, with one formed by the heavy quasiparticles, condensing into Cooper pairs below $T_c$, and the second associated with the local antiferromagnetism,
which is unaffected in the superconducting state (see for example \cite{caspary}) and characterized by rather large moments. Finally, at least two heavy-fermion superconductors, UPt$_3$ and URu$_2$Si$_2$, appear to exhibit an apparent coexistence between weak antiferromagnetism (WAF),  {\it i.e.} with a ordered moment $\mu_s \ll 0.1~\mu_B/U$, occurring at rather high temperature and superconductivity appearing at a fraction of the N\'eel temperature \cite{aeppli,schlabitz}.

The purpose of the present article is to review recent $\mu$SR studies on the ground state properties of these latter heavy-fermion compounds, particularly as function of control parameters like chemical substitution and external pressure (see also \cite{devisser,keizer,graf,amitsuka,amitsuka_2}).

\section{Particularities of the $\mu$SR technique}

$\mu$SR provides a number of advantages for investigating the ground-state properties of the heavy-fermion compounds (see for example \cite{amato}).

\begin{figure}
\begin{center}
\includegraphics[scale=0.4]{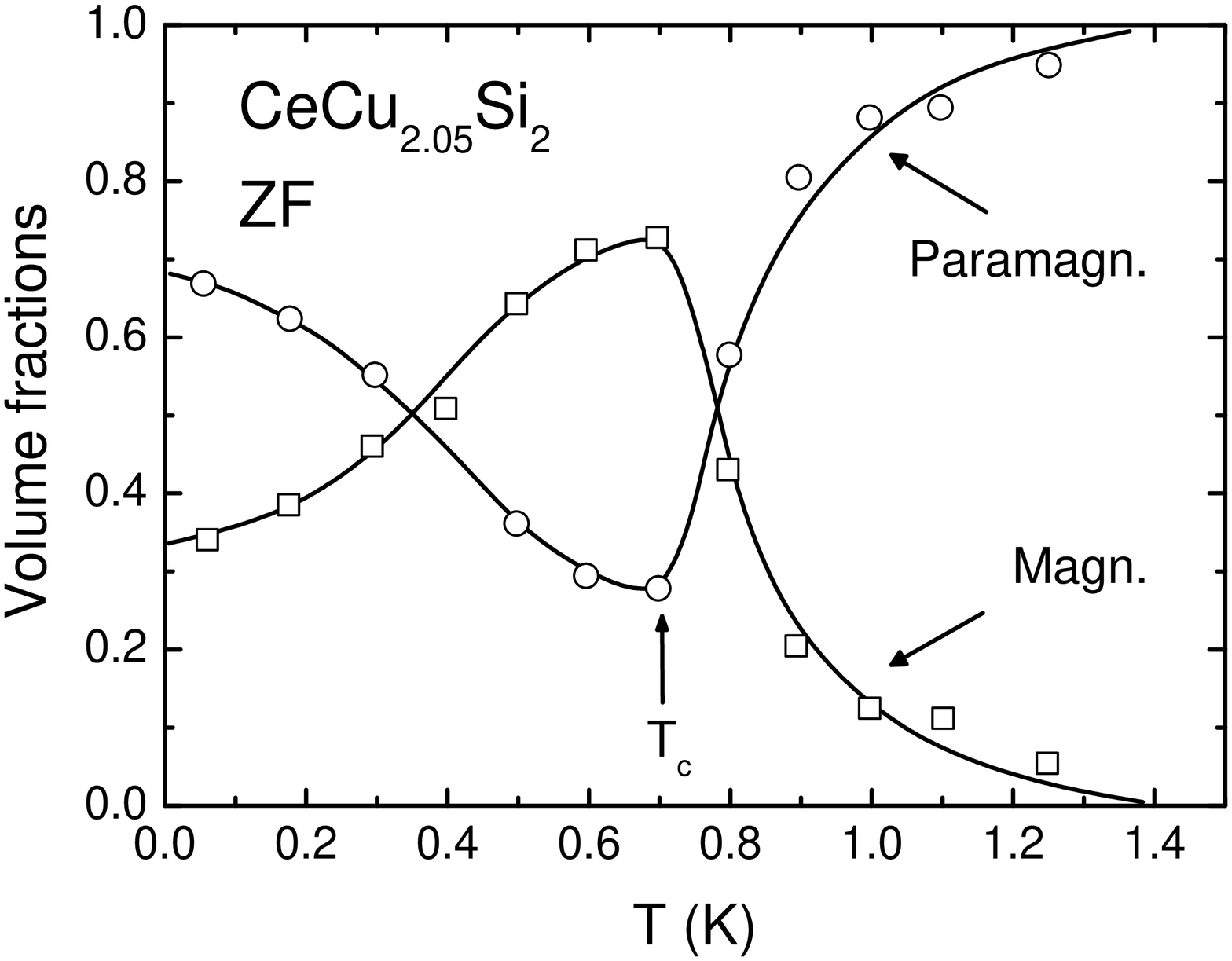}
\caption{\label{cecu2si2} Temperature dependence of the magnetic and paramagnetic volume fractions in CeCu$_2$Si$_2$ obtained by zero-field $\mu$SR. Note the {\it decrease} of the magnetic fraction upon cooling the sample below the superconducting temperature (adapted from \cite{feyerherm}).}
\end{center}
\end{figure}

The $\mu$SR technique is sensitive to extremely small internal fields (down to $\sim$0.1~G) and therefore can probe local magnetic fields that may be nuclear or electronic in origin. Since no applied field is necessary to polarize the spin of the implanted muons, such measurements can be made in the absence of any perturbative external field. 
$\mu$SR has also the capability to detect temporal as well as spatial changes of the internal fields. In this respect, the local-probe character of the muon makes $\mu$SR very sensitive to spatially inhomogeneous magnetic properties. Hence the occurrence of different phases in a sample will be reflected by different components in the $\mu$SR signal, and a careful analysis of these components furnishes a direct measure of the fraction of the sample volume involved in a particular phase. For heavy-fermion compounds, the $\mu$SR technique can therefore be utilized to check the real coexistence of different types of ground states at the microscopic level that is only assumed by traditional macroscopic techniques. An exemplary study demonstrating the capability of the technique is given by the measurements on the heavy-fermion CeCu$_2$Si$_2$, which exhibits heavy-fermion superconductivity and/or a weak magnetic order which are extremely sensitive to the exact stoichiometry of the samples \cite{grewe}. In addition to the first determination of the magnetic character of the phase transition at $\sim$1~K \cite{uemura}, the $\mu$SR technique has also clearly demonstrated that this latter phase does {\it not} coexist but rather competes for volume with the superconducting phase \cite{luke,feyerherm}. This conclusion is based on the temperature evolution of the amplitudes of the clear two-component $\mu$SR signal at low temperature, furnishing a direct measure of the volume fractions related to both phases (see \fref{cecu2si2}) \cite{stockert}.

\section{UPt$_3$}
The low-temperature normal state properties of the hexagonal heavy-fermion material UPt$_3$ are exemplary of a strongly renormalized Fermi-liquid system with a quasiparticle mass of the order of 200 times the free electron value \cite{stewart}. UPt$_3$ appears to be the 
heavy-fermion compound for which the most indications for unconventional superconductivity have been reported (for a review see \cite{taillefer}). Careful measurements on high-quality crystals reveal the presence of a double superconducting transition (at about 0.5~K \cite{fisher}) in zero external field and a striking multicomponent diagram with three superconducting phases meeting at a tetracritical point in applied magnetic fields \cite{bruls}. In addition, peculiar magnetic properties are observed, such as the occurrence of so-called small-moment antiferromagnetic order (SMAF) which is found below $T_{\rm N, SMAF} \simeq 6$~K \cite{aeppli_2}. The ordered phase, which is characterized by an unusually small magnetic moment ($\mu_s=0.02$~$\mu_{\rm B}$/U directed along the $a^*$-axis) was solely observed by neutron diffraction and magnetic x-ray scattering \cite{aeppli_2,hayden,isaacs}. Thermodynamic and transport properties studies, as well as microscopic studies (NMR \cite{tou} and $\mu$SR \cite{keizer,dalmas}), could not reliably detect the SMAF order.

\begin{figure}
\begin{center}
\includegraphics[scale=0.4]{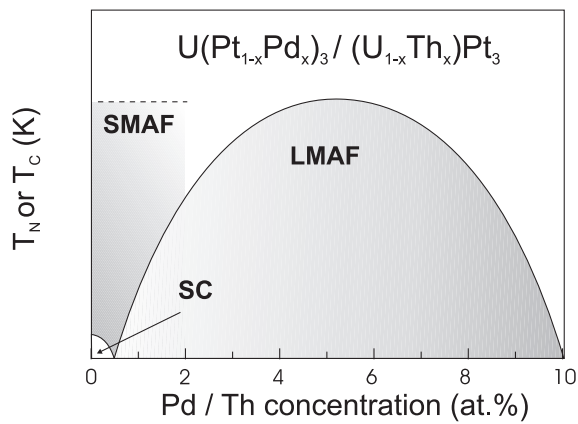}
\caption{\label{upt3} Schematic magnetic and superconducting phase diagram for U$_{1-x}$Th$_x$Pt$_3$ and U(Pt$_{1-x}$Pd$_x$)$_3$ alloys. Whereas the SMAF state, occurring below $T_{\rm N,SMAF}\simeq 6$~K, is only observed by neutron diffraction and magnetic x-ray scattering, the LMAF and the superconducting (SC) states are detected by different techniques.}
\end{center}
\end{figure}

Besides the SMAF order, conventional antiferromagnetism ({\it i.e.} with rather large moments) can be induced upon substituting small amounts of Pd or Au for Pt \cite{batlogg, devisser_2}, or when U is replaced by small amounts of Th \cite{ramirez}. For both pseudo-binary systems U$_{1-x}$Th$_x$Pt$_3$ and U(Pt$_{1-x}$Pd$_x$)$_3$ the so-called large-moment antiferromagnetism (LMAF) can be induced in the concentration range of about $0.01\leq x \leq 0.1$, with a maximum for the N\'eel temperature of $T_{\rm N,max, LMAF} \simeq 6$~K at about 5 at.\% Th or Pt (see \fref{upt3}  and \cite{frings, keizer_2}). The close resemblance of the phase diagram of both pseudo-binary systems indicates that the localization of the uranium moments is governed by the $c/a$ ratio of the lattice parameters which decreases for both dopants \cite{batlogg,devisser_3,opeil}. Although the LMAF state is characterized by magnetic moments at least one order of magnitude stronger that those observed in the SMAF phase, the magnetic structure is identical for both phases. Moreover, the transition temperature of the SMAF phase and the highest transition temperature of the LMAF phase appear strikingly equal, pointing to a close interplay between these phases. 

In UPt$_3$ the occurrence of multiple superconducting phases is usually accounted for by phenomenological Ginzburg-Landau theories involving complex two-dimensional superconducting order parameters, whose degeneracies are lifted by a symmetry-breaking field (SBF), usually associated to the SMAF order \cite{taillefer}. While it is generally accepted that the Cooper pairing is due to magnetic correlations, the odd parity of the order parameter, demonstrated by different techniques \cite{tou, dalichaouch}, appears at first glance incompatible with a picture where the dominant spin fluctuations are of antiferromagnetic nature. Hence, as noticed by several authors at an early stage of the heavy-fermion research ({\it e.g.} Anderson \cite{anderson} or Miyake \etal \cite{miyake}), in simple models odd-parity pairing is favorised by ferromagnetic fluctuations. This long standing controversy has triggered comprehensive $\mu$SR studies on the pseudo-binaries U(Pt$_{1-x}$Pd$_x$)$_3$ and U$_{1-x}$Th$_x$Pt$_3$. 

Let us first concentrate on the $\mu$SR results obtained for the U(Pt$_{1-x}$Pd$_x$)$_3$ pseudo-binary system \cite{keizer}. A great deal of attention has been given to the nature of the SMAF state. $\mu$SR studies performed on high quality UPt$_3$ crystals \cite{dalmas}, as well as NMR studies \cite{tou}, lead to the conclusion that this state does not involve static moments but rather possesses moments which fluctuate at a rate larger than 100~MHz, yet slower than the nearly-instantaneous scales of neutron and x-ray scattering. 

\begin{figure}
\begin{center}
\includegraphics[scale=0.4]{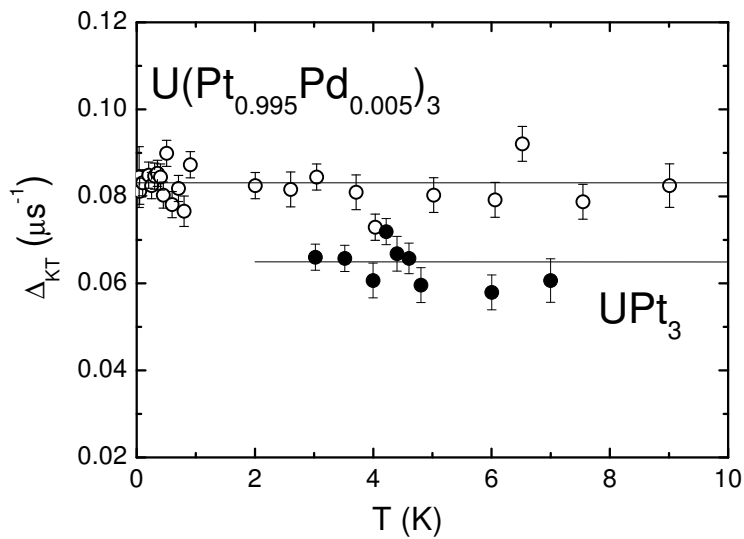}
\caption{\label{upt3_zf} Temperature dependence of the zero-field $\mu^+$-depolarization rate obtained on high quality samples of UPt$_3$ and U(Pt$_{0.995}$Pd$_{0.005}$)$_3$. Note the absence of any anomaly in the vicinity of $T_{\rm N,SMAF}\simeq6$~K. The difference in absolute values between both measurements is marginal and arises mainly from the slight change of the lattice constants}
\end{center}
\end{figure}

\begin{figure}
\begin{center}
\includegraphics[scale=0.4]{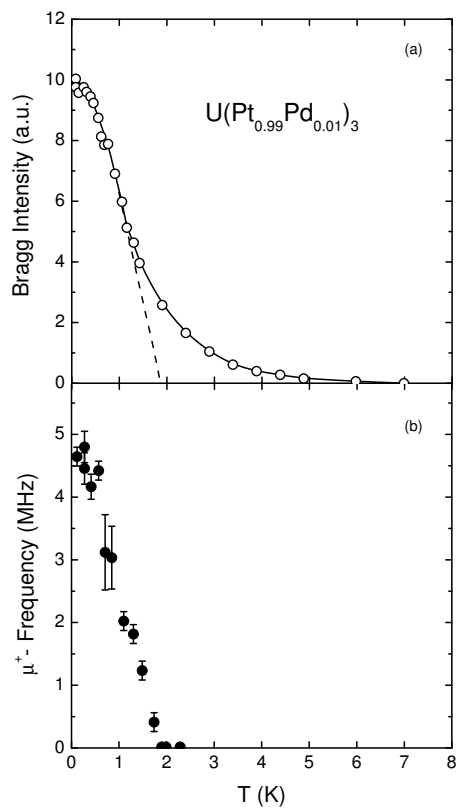}
\caption{\label{upt3_neutron_muon} (a) Temperature variation of the magnetic intensity measured at the magnetic Bragg peak ${\mathbf Q}=(0.5,0,1)$ for an annealed U(Pt$_{0.99}$Pd$_{0.01}$)$_3$ sample. The sharp increase in the intensity near 1.9~K indicates
a crossover from SMAF to LMAF. (b) Spontaneous $\mu^+$-frequency obtained on the same sample. Note the absence of magnetic signal above $T_{\rm N,LMAF}\simeq 1.9$~K.}
\end{center}
\end{figure}

For $\mu$SR experiments, this conclusion was inferred from the absence of any anomaly of the muon depolarization rate when cooling the sample into the SMAF phase. The alternative explanation, for the absence of a SMAF signature in the $\mu$SR signal, invoking the cancellation of the internal fields at the muon stopping site in the SMAF phase, could be safely discarded since it does not hold for the Pt site and is therefore incompatible with the NMR results.
The $\mu$SR results, indicating the dynamic nature of the SMAF phase, were obtained not only on pure UPt$_3$ samples, but also on samples with low ({\it i.e.} $x \le 0.005$) Pd concentration (see \fref{upt3_zf}). $\mu$SR studies on a U(Pt$_{0.99}$Pd$_{0.01}$)$_3$ sample \cite{keizer}, which according to neutron scattering studies should exhibit both magnetic phases, could only detect the LMAF phase (see \fref{upt3_neutron_muon}). Therefore, the $T_{\rm N,SMAF}$ should be considered as a crossover temperature signaling a slowing down of magnetic fluctuations rather than being regarded as a true phase transition temperature. Note that an analogous picture has been recently deduced, comparing $\mu$SR and neutron studies, for the high-temperature superconductor YBa$_2$Cu$_3$O$_{6.5}$ \cite{sidis}.

For $T > T_{\rm N,LMAF}$, the muon spin depolarization of the U(Pt$_{1-x}$Pd$_x$)$_3$ samples is found to result from the Gaussian distribution of static, 
randomly-oriented, magnetic fields due to $^{195}$Pt nuclei. As expected, the form of the depolarization function is given by the 
Kubo-Toyabe function
\begin{equation}	
\label{equation_kt}
  {\cal A}_{\rm{KT}}\,G_{\rm{KT}}(t) = {\cal A}_{\rm{KT}}\Bigg[\frac{1}{3} +
  \frac{2}{3}\Big(1-\Delta_{\rm{KT}}^2t^2\Big)\exp\Big(-\frac{1}{2}\Delta_{\rm{KT}}^2t^2\Big)\Bigg]~.
\end{equation}
Since there is no zero-field $\mu$SR signature for the SMAF state equation\,\eref{equation_kt} works equally well in the paramagnetic phase as in the anomalous SMAF region. On the other hand, the signature of the LMAF phase is characterized by the occurrence of a two-component function in the $\mu$SR signal:
\begin{eqnarray}
\label{equation_osc}
{\cal A}_{\rm{LMAF}}\,G(t)_{\rm{LMAF}}&=&{\cal A}_{\rm{osc}}\Bigg[\,\frac{2}{3}\exp(-\lambda t)\cos(2\pi\nu t+\phi)+\frac{1}{3}\exp(-\lambda't)\,\Bigg]\nonumber\\ 
&&+{\cal A}_{\rm{KL}}G_{\rm{KL}}(\lambda_{\rm{KL}},t)~.
\end{eqnarray}
These two components are indicative of two magnetically inequivalent muon stopping sites. Whereas one site presents a finite local field $B_{\mu}$ (first term on the right hand side of equation\,\eref{equation_osc} characterized by the frequency $\nu=\gamma_{\mu}B_{\mu}/(2\pi)$ (where $\gamma_{\mu}$ is the gyromagnetic ratio of the muon), the second site is characterized by an isotropic Lorentzian distribution of local fields with an average zero value producing a Kubo-Lorentzian polarization decay (second term on the right hand side of equation\,\eref{equation_osc}). Although such deductions could appear, at a first glance, of specific interest solely for $\mu$SR specialists, they may disclose some subtle crystallographic details. Additional transverse-field $\mu$SR measurements in the paramagnetic phase in pure and doped systems revealed the presence of two magnetically distinct muon stopping sites \cite{schenck}. However, a careful analysis of the $\mu^+$-Knight shift as a function of a modified Curie-Weiss magnetic susceptibility indicate that both signals have the same crystallographic origin, namely the $2a$ site (0,0,0), but are characterized by different contact-field contributions. This difference can be attributed to slightly different electron density and/or by a dissimilar polarizability of these electrons at the muon stopping site. 

Such observation provides strong evidence for two distinct regions of different magnetic response in the magnetic as well as in the paramagnetic phases. However, the $\mu$SR data alone do not allow one to distinguish whether the different response arises from macroscopically separated domains or from a crystallographic modulation. Interestingly, several reports have been published on structural modulations in UPt$_3$ detected by transmission electron microscopy (TEM). Midgley \etal \cite{midgley} reported a complex set of incommensurate structural modulations at room temperature, corresponding to several ${\mathbf q}$-vectors of magnitude around $0.1\pi/a$. Studies performed on a whisker by Ellman \etal \cite{ellman} found also a well-developed incommensurate modulation of the same magnitude with a single ${\mathbf q}=(0.1,-0.1,-0.1)$, corresponding to a modulation of wavelength $\sim70$~\AA. Finally, we mention an x-ray and TEM study by Aronson \etal \cite{aronson} suggesting that stacking faults could play a predominant role for intrinsic disorder in UPt$_3$. The stacking sequence becomes ABACABAC, corresponding to a double hexagonal structure. However, in this latter scenario, the atoms in the A layer of the double hexagonal sequence possess an environment with cubic symmetry, and therefore should have an isotropic susceptibility which was not detected by angular scans of the $\mu^+$-Knight shift. Finally, for sake of completeness, we mention an x-ray study reporting the observation of a weak trigonal distortion (space group $P\bar{3}m1$) of the hexagonal structure \cite{walko}. This observation was proposed as possible explanation for the different signals observed in the $\mu$SR spectra since the 2a sites (0,0,0) and (0,0,0.5) are no longer equivalent in the tetragonal lattice. However, recent high resolution x-ray studies, in particular Renninger scans around the forbidden reflections, may rule out the possibility of a trigonal distortion \cite{yaouanc-private}.

The idea to associate a structural modulation with the SBF, needed by a number of theories to explain the occurrence of multiple superconducting phases, has been put forward by several theorists \cite{machida, mineev}. As a matter of fact, some observations appear to be incompatible with a magnetically driven SBF. The status of SMAF state as a true long-range order parameter is itself in doubt  since, in addition to its dynamical character, its correlation length as determined by neutron diffraction is at most 300 \AA, {\it i.e.} of the order of the superconducting coherence length. This has raised the relevent criticism that the SMAF is unable to break the hexagonal symmetry. Other questions are raised by neutron studies under pressure \cite{hayden, hilbert}. 
The close correlation between the splitting of the superconducting transition and the decrease of the magnetic moment, which is taken as strong evidence of a magnetically mediated SBF, is difficult to reconcile with a N\'eel temperature which appears to rise slightly under pressure, implying that $T_{\rm N,SMAF}(p)$ vanishes in the middle of the $p-T$ plane, which is forbidden \cite{chen}.

Besides its potential role for theories involving an order parameter belonging to a two-dimensional representation of the hexagonal point group, the question arises whether a crystallographic modulation, as possibly detected by muons and involving subtle changes of the density and/or polarizability of the conduction electrons, could somehow fit into theories involving two independent, one-dimensional, nearly degenerate superconducting order parameters \cite{taillefer}. In such theories, the splitting of the superconducting phase transition is due to accidental degeneracy, not to coupling to a symmetry-breaking field. This is exemplified by studies on the so-called AB-model, consisting of like-parity order parameters, one transforming as an A and the other as a B representation of the hexagonal point-group D$_{6h}$ \cite{garg}. 

In view of the peculiar nature of the SMAF state characterized by small moments and a dynamical character, it appeared also quite important to investigate the relation between the static magnetic order (i.e. LMAF) and superconductivity. Indeed, and as said above, a characteristic feature of heavy-fermion superconductivity is the proximity to a magnetic quantum critical point, corresponding to a phase transition at $T =0$. Several pressure studies on antiferromagnetically ordered systems have demonstrated that the approach to the magnetic quantum critical point is connected to the occurrence of superconductivity \cite{mathur, saxena}. In addition, the superconducting transition temperature $T_c$ is found to have a maximum at the critical pressure for which static antiferromagnetism disappears and strong spin fluctuations are present. This is interpreted as evidence that superconductivity is mediated by antiferromagnetic fluctuations.

\begin{figure}
\begin{center}
\includegraphics[scale=0.4]{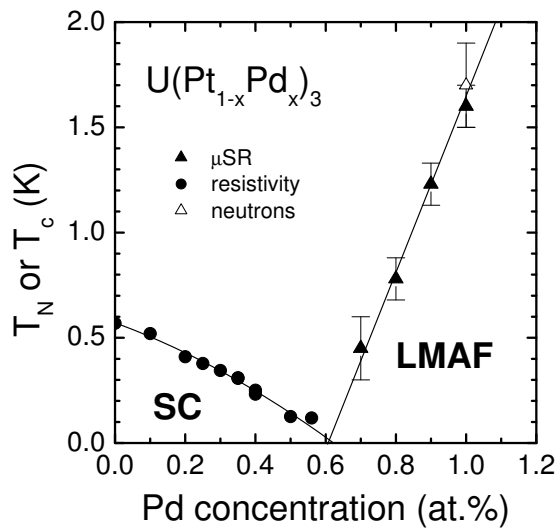}
\caption{\label{upt3_precise_phase_diagram} Magnetic and superconducting phase diagram for (UPt$_{1-x}$Pd$_x$)$_3$ alloys with $x < 0.012$. The superconductivity phase boundary was obtained by resistivity. The solid lines serve to guide the eye (adapted from \cite{devisser}).}
\end{center}
\end{figure}

\Fref{upt3_precise_phase_diagram} represents the superconducting and magnetic phase diagram of U(Pt$_{1-x}$Pd$_x$)$_3$ near the quantum critical point of the LMAF state. Due to the extremely small ordered magnetic moments, the LMAF phase in the vicinity of the quantum critical point could only be observed by $\mu$SR technique. The Pd-concentration dependence of the superconducting transition temperature was recorded by electrical resistivity. The main point is that the LMAF phase represents indeed the magnetic instability in the pseudo-binary system and that the critical concentration for the suppression of superconductivity coincides with the critical concentration for the appearance of static antiferromagnetism, i.e. $T_c \rightarrow 0$ at the magnetic quantum critical point. This observation appears to rule out a superconducting singlet state mediated by antiferromagnetic fluctuations, as a maximum of $T_c$ should be expected as $T_{\rm N,LMAF} \rightarrow 0$, and seems to favorise a picture of superconducting triplet state. In UPt$_3$, evidence for odd-parity pairing is provided by a number of other experimental findings, such as the temperature-independent NMR \cite{tou} and $\mu$SR Knight-shifts below $T_c$ \cite{luke_2}. In the same vein, studies of the depression of the superconducting transition temperature as function of impurity doping strongly suggests the leading role of potential scattering and the absence of spin-flip scattering, pointing therefore to an odd-parity of the order parameter \cite{dalichaouch}. 

Assuming a superconducting triplet state, the coincidence between the suppression of $T_c$ and the magnetic quantum critical point could originate from pairbreaking process arising from the reduction of quasiparticle coherence and lifetime due to scattering by spin fluctuations \cite{wang}. Alternatively, the observed phase diagram may suggest that ferromagnetic fluctuations play a relevant role. Hence, besides the presence of antiferromagnetic fluctuations, several studies point to the existence of ferromagnetic fluctuations in UPt$_3$. Indications of ferromagnetic fluctuations came first from specific heat studies \cite{stewart} where a $T^3\ln (T/T_{SF})$-contribution similar to the one detected in $^3$He, and explained by Doniach and Engelsberg in terms of ferromagnetic fluctuations \cite{doniach}, is detected. Inelastic neutron studies indicate also the presence of fluctuations centered at ${\mathbf q}=0$ and accounting for about 20\% of the total susceptibility \cite{bernhoeft, goldman}. Consequently, the phase diagram of U(Pt$_{1-x}$Pd$_x$)$_3$ near the critical concentration might be understood by assuming that the superconducting state is mediated by ferromagnetic fluctuations and that the Pd-doping lead to a shift of spectral weight from ferromagnetic fluctuations towards antiferromagnetic fluctuations \cite{devisser}. A dominant role played by ferromagnetic fluctuations in mediating the non-conventional superconducting state in UPt$_3$ appears compatible with models \cite{anderson,miyake,fay}, for which odd-parity pairing is favorised by ferromagnetic fluctuations. 

We now turn to the results obtained on the U$_{1-x}$Th$_x$Pt$_3$ series which, as described above, presents a close analogy to the previously described U(Pt$_{1-x}$Pd$_x$)$_3$ systems \cite{graf}. As for the pure case, experiments performed on a polycrystalline $x=0.002$ sample indicate that the muon depolarization is well-described by a simple Kubo-Toyabe function (see equation\,\eref{equation_kt}) arising solely from the $^{195}$Pt nuclear moments. Therefore, the $\mu$SR data, here again, point to the dynamical nature of the SMAF state.

\begin{figure}
\begin{center}
\includegraphics[scale=0.4]{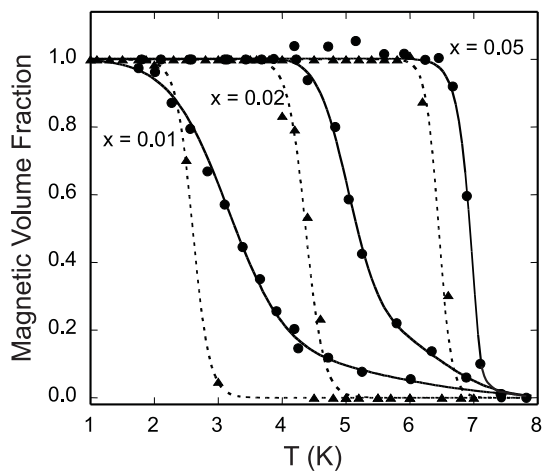}
\caption{\label{upt3_transitions} Comparison between the temperature dependent magnetic volume fractions of U$_{1-x}$Th$_x$Pt$_3$ ($\fullcircle$) and U(Pt$_{1-x}$Pd$_x$)$_3$ ($\blacktriangle$) alloys. Solid and dashed lines are guides to the eye. Note the broad transition width for the Th-substituted data.}
\end{center}
\end{figure}

The most striking aspect of the data obtained for Th concentrations $x>0.006$ is that the $\mu$SR signal is best described by a sum of equations\,\eref{equation_kt} and \eref{equation_osc} over an anomalous broad temperature range around the mean-field values of $T_{\rm N,LMAF}$. \Fref{upt3_transitions} shows the temperature evolution of the normalized magnetic fraction of the total amplitude of the $\mu$SR signal, {\it i.e.} ${\cal A}_{\rm{LMAF}}/({\cal A}_{\rm{LMAF}}+{\cal A}_{\rm{KT}})$, for Th-substituted samples with $x=0.01$, 0.02 and 0.05. For comparison, the same quantity for the corresponding Pd-substituted samples is reported. Whereas all the Pd-substituted samples have narrow transition widths, only the $x=0.05$ Th-substituted sample does. On the other hand, the $x=0.01$ and 0.02 Th-substituted samples exhibit rather large transition widths extending up to 7~K, which corresponds to the transition temperature of the $x=0.05$ sample. As discussed in Ref.~\cite{graf}, chemical inhomogeneities 
form an unlikely cause for the observed broadening. 
The broad transitions might possibly find their origin in a slowing down of the fluctuations of the SMAF state upon Th doping, in a way that the fluctuation timescale becomes comparable to the typical $\mu$SR timescale. This scenario is strongly supported by the observation that the magnetic $\mu$SR signal starts to rise at a fixed temperature ({\it i.e.} $\simeq7$~K) independent of the Th-concentration. Indeed and as demonstrated by neutron diffraction studies for U(Pt$_{1-x}$Pd$_x$)$_3$ \cite{keizer_2} the SMAF state is shown to be quite robust  upon alloying ({\it i.e.} $T_{\rm N,SMAF}$ independent of $x$) and a similar situation is expected for the Th-substituted analogue.

The idea that crystallographic disorder could somehow affect the fluctuation timescale of the SMAF state sheds a new light on very early $\mu$SR data on UPt$_3$. Cooke \etal \cite{cooke} observed for a polycrystalline sample a clear rise of the depolarization below 5~K, owing to the occurrence of static magnetic moments of electronic origin. As stated above, such observation could not be confirmed by subsequent $\mu$SR studies on high-quality single crystals, suggesting that sample quality may play a key role on the SMAF fluctuation timescale.

Although some aspects of the WAF phenomenon in UPt$_3$ remain to be solved, this system constitutes a striking example wherein the specificity and high sensitivity of the $\mu$SR technique has uncovered a number of unusual features, shedding new light on the interplay between magnetism and superconductivity in heavy-fermion superconductors.

\section{URu$_2$Si$_2$}
The heavy-fermion compound URu$_2$Si$_2$ exhibits two successive phase transitions at 17.5 and 1.4~K. Whereas, the transition at 1.4~K signals the occurrence of unconventional superconductivity \cite{schlabitz}, the phase transition at $T_{\rm o}\simeq 17.5$~K still remains mysterious. Below $T_{\rm o}$, the occurrence of a simple antiferromagnetic ordering with 5$f$ magnetic moments along the $c$-axis is suggested by neutron diffraction studies \cite{broholm}. The observed magnetic Bragg peaks are extremely weak indicating a static moment of 0.03~$\mu_{\rm B}$/U with ordering vector ${\mathbf Q}=(0,0,1)$. Such a deduction is hard to reconcile with large anomalies observed at $T_{\rm o}$ in the thermodynamical properties. In addition, the magnitude of the ordered moment reported in neutron- and x-ray magnetic-scattering measurements varies significantly between experiments (ranging from 0.017 up to 0.04~$\mu_{\rm B}$/U) and cannot be simply ascribed to experimental uncertainties. Similar discrepancies between studies are reported for the temperature dependence of the ordered moment. Consequently, different scenarios have been invoked concerning the true nature of the phase below $T_{\rm o}$. Several models suggest that the observed macroscopic anomalies are not connected to the magnetic state reported by neutron studies, but rather should be associated with a hidden order parameter \cite{barzykin,ikeda,santini,ohkawa,tsuruta}. For example, the possibility of a quadrupolar order has been invoked, but to date no compelling experimental evidence could be reported.

An additional controversy has recently appeared when comparing different microscopic measurements performed under external pressure. On one side, neutron scattering measurements \cite{amitsuka_3,amitsuka_4} revealed that the magnetic Bragg-scattering intensity is significantly enlarged with pressure. This was interpreted as an increase of the staggered moment up to $\mu_s\simeq 0.25~\mu_{\rm B}$/U at about 1~GPa. On the other side, $^{29}$Si-NMR measurements \cite{matsuda} suggest that solely the magnetic volume fraction is affected under pressure, but that the staggered moment remains essentially {\it constant}. This was deduced from the fact that the central paramagnetic NMR line is split into two symmetrically located lines below $T_{\rm o}$, whose positions remain constant, whereas their intensities increase with pressure. Hence, the weak magnetic signal observed by neutron studies below $T_{\rm o}$ at ambient pressure should not be connected to a small staggered moment but rather to a vanishingly small magnetic volume fraction of the order of 1\% (corresponding to the ratio of the Bragg intensities between 0 and 1~GPa {\it i.e.} $\simeq (0.03/0.25)^2$) \cite{amitsuka_3}. Such a deduction could also shed a new light on the anomalous short-range correlation length $\xi$ of the weak magnetism determined by neutron scattering (see for example \cite{broholm}). The reported values ($\xi\simeq 200-400$~\AA) appear too short for a stable magnetic state but could be reasonable if $\xi$ represents in fact a measure of the size of the magnetic clusters.

In view of its ability to detect different magnetic responses, $\mu$SR under pressure was used to provide an additional microscopic point of view \cite{amitsuka,amitsuka_2}. The main goal was to gain more insight on a possible interplay between the hidden order state and the peculiar antiferromagnetic state.

Let first discuss the main findings obtained by $\mu$SR measurements at ambient pressure. The majority of the studies indicate a marginal increase of the muon depolarization rate reflecting a slight and isotropic increase of the field distribution at the muon stopping site \cite{amitsuka_4,knecht,maclaughlin,yaouanc}. The internal field width at the muon site is of the order of 0.1~G. Such a value is orders of magnitude smaller that the one calculated for the majority of possible $\mu^+$ stopping sites considering the weak antiferromagnetic state suggested by neutron scattering. Consequently, and to reconcile neutron and $\mu$SR studies, in some studies the muon was assumed to stop at the $f$ site (\small{$\frac{1}{4}\,\frac{1}{4}\,\frac{1}{4}$}\normalsize), which is symmetric between the two magnetic sublattices and for which the dipolar fields cancel (see for example \cite{maclaughlin}). However, $\mu$SR measurements performed on a number of isostructural compounds appear to rule out any occupation of the $f$ site by the muon in the ThCr$_2$Si$_2$-type structure \cite{amato_2,dalmas_2,nojiri}. Therefore, it emerges that the observed increased field-distribution is somehow connected to the hidden order state and not to a strongly reduced magnetic volume. This appears in line with NMR studies reported by Bernal \etal \cite{bernal} showing that at $T<T_{\rm o}$ the central ({\it i.e.} the paramagnetic non-split) silicon NMR line displays an isotropic increase of its width of the order of few Gauss. Here again, this field magnitude is too small to be explained by the moment deduced by neutron scattering. Moreover, a cancellation of the dipolar fields is excluded at the silicon site and, furthermore, the moment deduced by neutron diffraction is aligned along the $c$-axis and thus cannot account for the isotropic nature of the local field distribution detected by NMR or $\mu$SR. In other words, this means that the hidden order state is characterized by an isotropic field distribution and not by the weak magnetic signal observed by neutron diffraction, which was shown to arise from a strongly reduced volume fraction. Note that the presence of a minority magnetic phase at ambient pressure was also detected by $\mu$SR on a particular single crystal \cite{luke_3}. The key point of this $\mu$SR study was the observation of a clear two-component structure of the $\mu$SR signal below $T_{\rm o}$. Whereas the predominant component (90\% of the total amplitude) exhibits a behavior similar to what was observed in other $\mu$SR studies ({\it i.e.}, weak increase of the depolarization rate), the second component shows clear oscillations, implying large internal fields in a fraction of the sample (spontaneous frequency of the order of $\nu_{\mu}\simeq 8.2$~MHz in about 10\% of the volume for $T\rightarrow 0$) created by an ordered moment of the order of 0.2~$\mu_{\rm B}$. This particular result confirms that the smallness of the magnetic Bragg-peak intensity in neutron-scattering experiments arose from a strongly sample-dependent reduced magnetic volume.

\begin{figure}
\begin{center}
\includegraphics[scale=0.4]{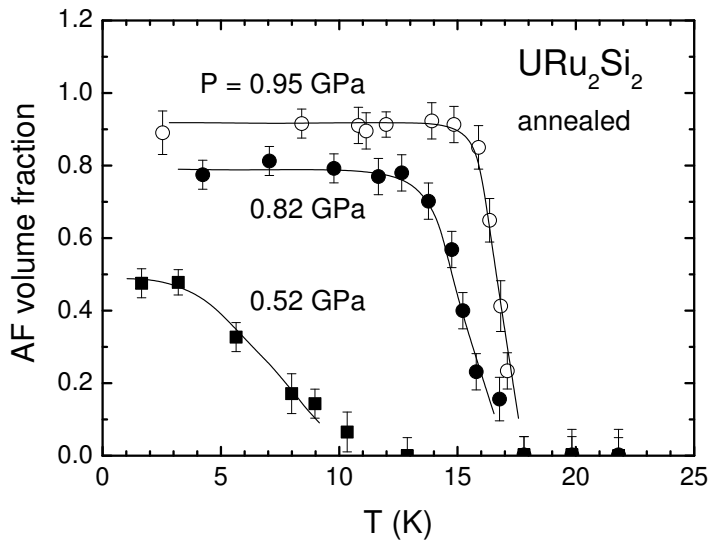}
\caption{\label{uru2si2_af_fraction} Temperature and pressure evolution of the antiferromagnetic fraction of URu$_2$Si$_2$ determined by $\mu$SR in an annealed single crystal. For clarity only few pressure curves are shown. Note that for low pressures the onset of the magnetic fraction ($T_{\rm M}$) is much lower than the hidden order temperature $T_{\rm o} \simeq 17.5$~K (see text).}
\end{center}
\end{figure}

\begin{figure}
\begin{center}
\includegraphics[scale=0.4]{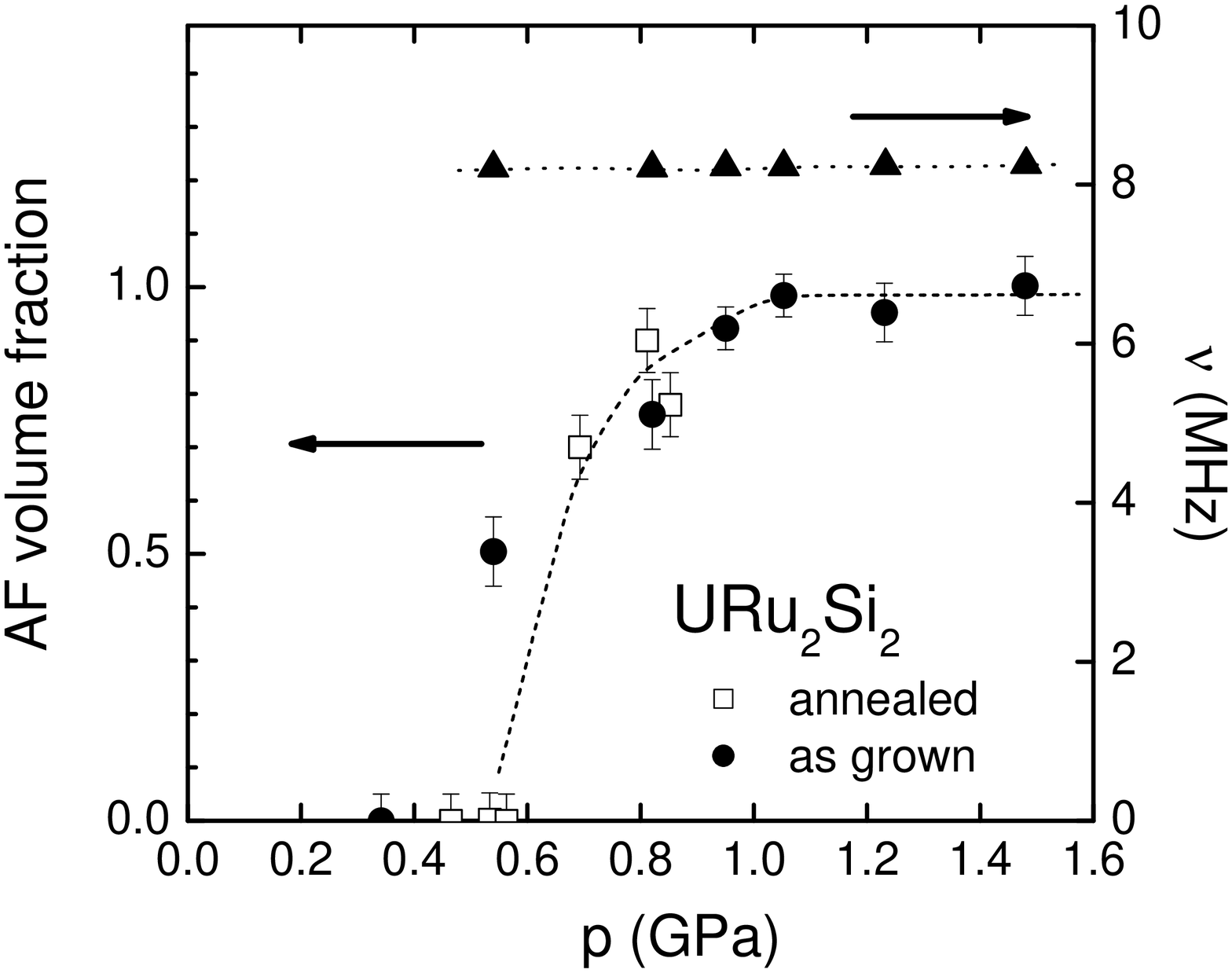}
\caption{\label{uru2si2_af_fraction_frequency} Pressure evolution of the antiferromagnetic fraction and spontaneous $\mu^+$-frequency, both extrapolated for $T\rightarrow 0$ on annealed and as-grown URu$_2$Si$_2$ single crystals.}
\end{center}
\end{figure}

$\mu$SR measurements under pressure were performed on high quality single crystals using a specially designed copper-beryllium (Cu-Be 25) clamped cell with a 1:1 mixture of $n$-pentane and isoamyl alcohol as transmitting medium \cite{andreica}. Pressures up to $p \simeq 1.5$~GPa could be reached. To pass the relatively thick cell walls, muons with rather high energy ({\it i.e.} $p_{\mu}\simeq 105$~MeV/c) are necessary and, therefore, measurements were performed on the $\mu$E1 beamline of the Paul Scherrer Institut (Villigen, Switzerland). The main finding is the observation of clear spontaneous oscillations in the $\mu$SR signal only for pressures above about 0.5~GPa (for a as-grown sample these pressures are shifted to about 0.7~GPa) \cite{amitsuka}. As shown on \fref{uru2si2_af_fraction}, the magnetic volume fraction is strongly pressure dependent and starts increasing at temperatures clearly lower than $T_{\rm o}$, as evidenced by the measurements performed at 0.52~GPa where the onset of the volume fraction is located at $T_{\rm M}\simeq11$~K. A full magnetic state is obtained for pressures of the order of about 1~GPa (see \fref{uru2si2_af_fraction_frequency}). The amplitude of the spontaneous frequency is strongly dependent on the sample orientation and disappears when the initial muon-spin polarization is parallel to the $c$-axis. This indicates that the local fields point along the $c$-direction, {\it i.e.} a situation drastically different than the isotropic one observed at ambient pressure. An interesting point is the almost pressure-independent saturation value of the spontaneous frequency $\nu_{\mu}(T=0,p)\simeq 8.25$~MHz. Note that this value is in good agreement with that observed by Luke \etal at ambient pressure on a minority phase of a particular sample (see above and \cite{luke_3}). 

\begin{figure}
\begin{center}
\includegraphics[scale=0.4]{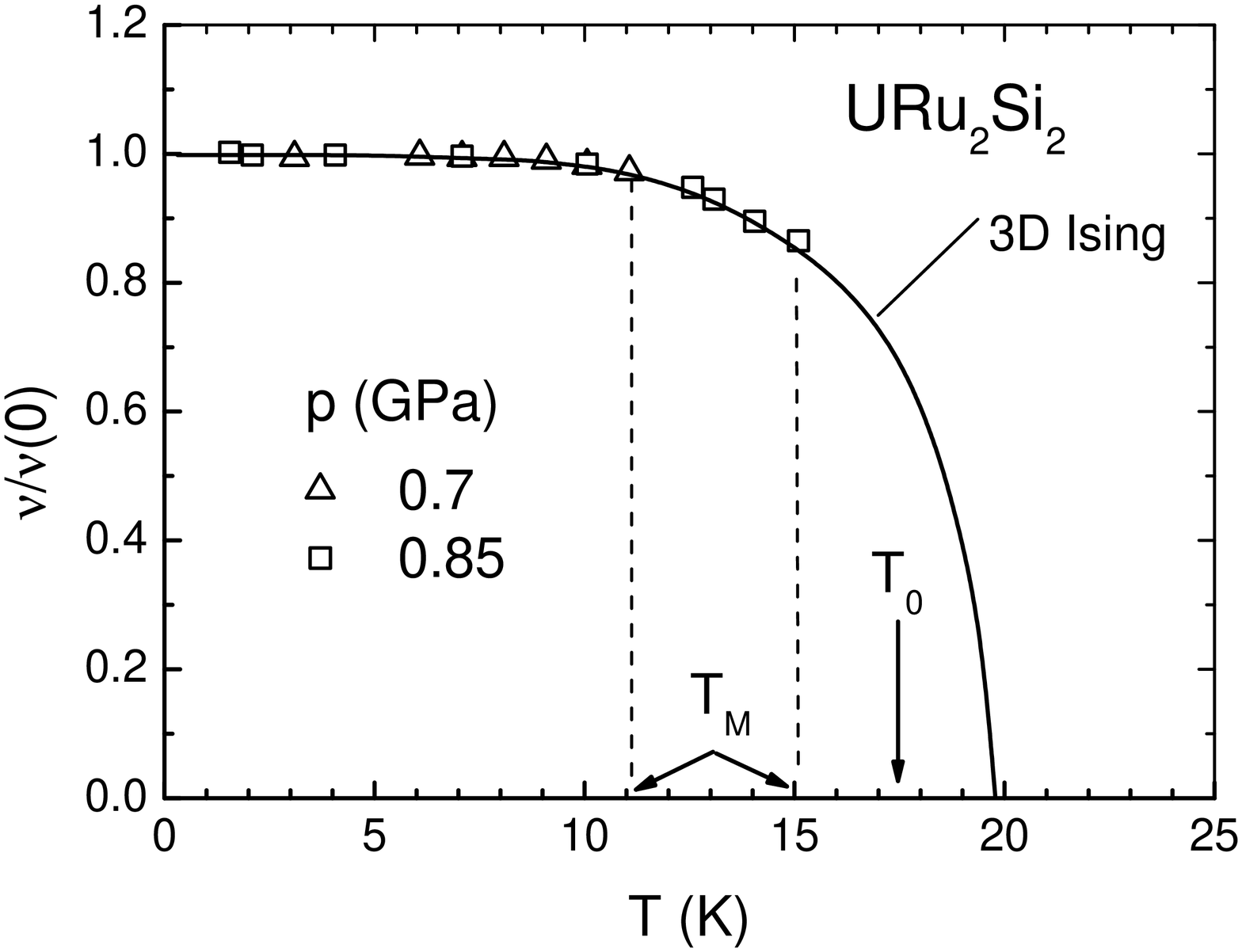}
\caption{\label{uru2si2_frequency} Temperature dependence of the spontaneous $\mu^+$-frequency for URu$_2$Si$_2$ under pressure. Note that all the data collapse on a universal curve well described by a 3D-Ising model. The different pressure measurements differ in the first-order transition temperature $T_{\rm M}$, which increases under pressure. Note that for clarity solely two different pressures are reported.}
\end{center}
\end{figure}

\begin{figure}
\begin{center}
\includegraphics[scale=0.6]{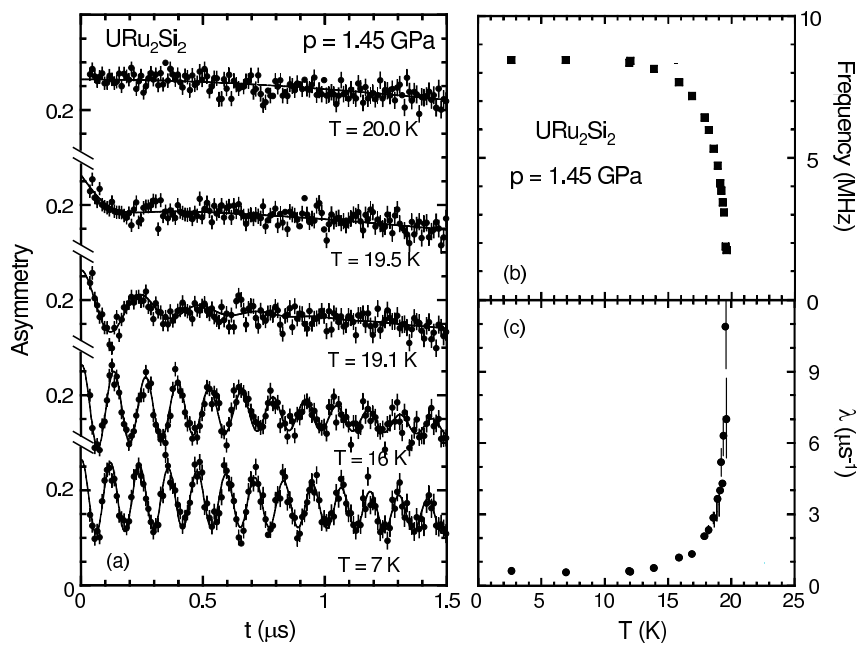}
\caption{\label{uru2si2_high_pressure} High pressure measurements ($p=1.45$~GPa) performed on an annealed URu$_2$Si$_2$ single crystal. For this pressure the system undergoes a direct second-order transition paramagnetism/antiferromagnetism ({\it i.e.} $T_{\rm M}\equiv T_{\rm N}>T_{\rm o}$). (a) Typical raw data. (b) Temperature evolution of the spontaneous frequency. (c) Transversal muon-depolarization rate exhibiting clear critical fluctuations when $T\rightarrow T_{\rm N}$.}
\end{center}
\end{figure}

The temperature dependence of the observed frequency, reported on \fref{uru2si2_frequency}, clearly indicates a first order transition between the hidden order state (occurring in the temperature range $T_{\rm M} \le T \le T_{\rm o}$) and the magnetic state present below $T_{\rm M}$. For all the measured pressures, the $\mu^+$-frequency data $\nu_{\mu}(T,p)$ collapse on a single curve, which is well described in terms of a 3D-Ising model. A fit of this model to the data leads to a nominal N\'eel temperature of $T_{\rm N}\simeq 20$~K, which, unlike $T_{\rm M}$, emerges therefore as pressure independent. Note that the value of $T_{\rm N}$ is higher than the hidden order temperature $T_{\rm o}$. By increasing the pressure, the onset temperature $T_{\rm M}$ can be shifted to values similar to the hidden order temperature $T_{\rm o}$. Once $T_{\rm M}>T_{\rm o}$  ({\it i.e.} for pressures $p>1$~GPa) the hidden order phase disappears and the system undergoes a direct transition paramagnetism/antiferromagnetism. For such pressures, a full temperature dependence curve for the ordered moment is observed, indicating a second order transition between the paramagnetic state and the antiferromagnetic state (see \fref{uru2si2_high_pressure}). Note that the observed $T_{\rm N}\simeq20$~K for $p>1$~GPa agrees quite well with the nominal $T_{\rm N}$, which was obtained by extrapolating the spontaneous-frequency curves at lower pressure. The occurrence of a first order transition between the hidden order phase and the antiferromagnetic phase was also deduced from very recent neutron scattering studies performed under pressure \cite{bourdarot}. However, such deduction was rather indirect and just relied on the increase of the temperature slope of the Bragg magnetic intensity under pressure. Due to the possibility to obtain independently a measure of the magnetic volume and of the magnetic moment, the present data provide on this context a much clearer observation. 

\begin{figure}
\begin{center}
\includegraphics[scale=0.6]{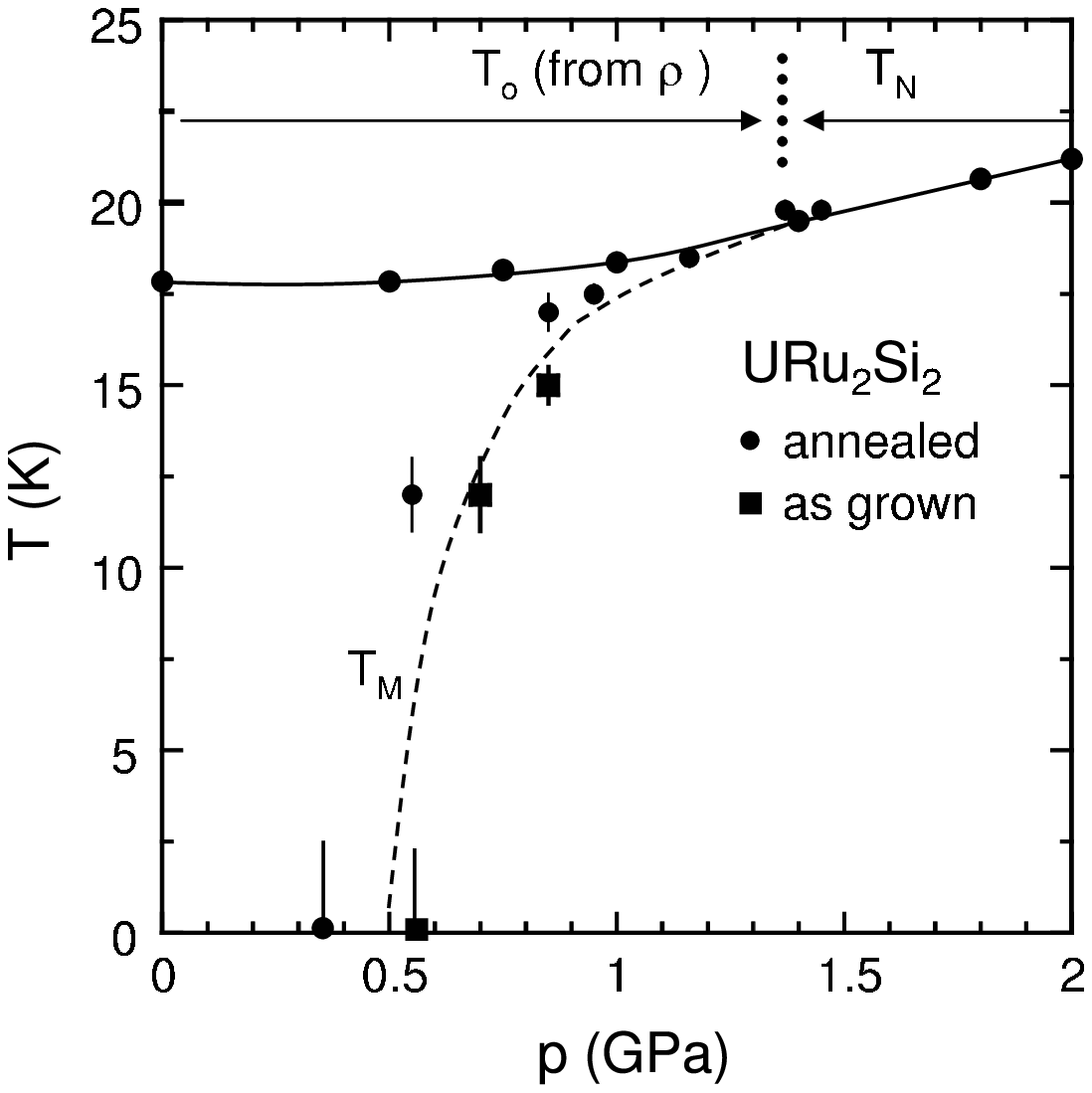}
\caption{\label{uru2si2_phase_diagram} $T-p$ phase diagram resulting from the present $\mu$SR data. The solid lines $T_{\rm o}(p)$ (determined by resistivity) and $T_{\rm N}(p)$ are second-order lines whereas $T_{\rm M}(p)$ is first-order. The arrows indicate regions with paramagnetism/hidden order transitions and paramagnetism/antiferromagnetism transitions.}
\end{center}
\end{figure}

\begin{figure}
\begin{center}
\includegraphics[scale=0.6]{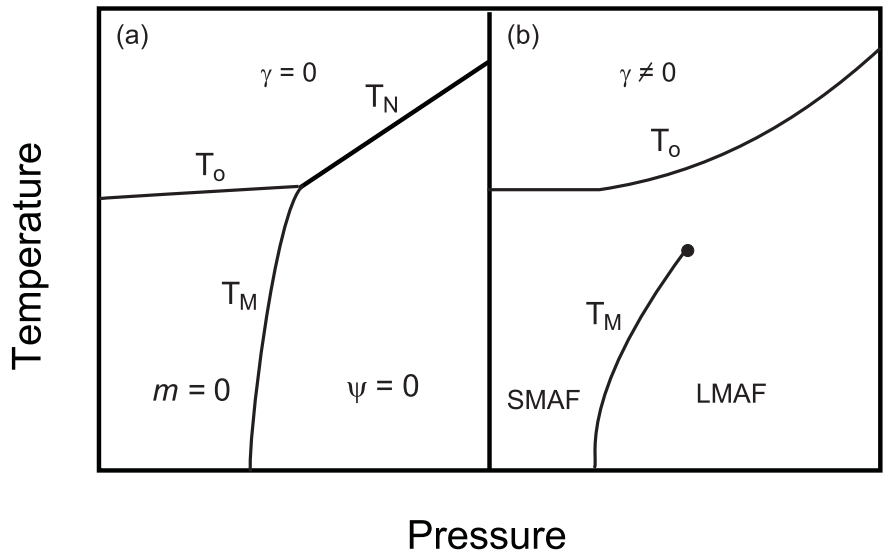}
\caption{\label{uru2si2_schematic} Schematic phase diagram derived from equation\,\eref{op_coupling} describing the coupling between the order parameters. (a) For $\gamma = 0$ and (b) for $\gamma \neq 0$ (adapted from \cite{bourdarot}).}
\end{center}
\end{figure}

The observations of (i) a first order line ($T_{\rm M}(p)$), between the hidden order phase and the magnetic phase; and (ii) its merging with second order lines delimitating the magnetic and the hidden order phases from the paramagnetic phase ($T_{\rm o}(p)$ and $T_{\rm N}(p)$ -- see \fref{uru2si2_phase_diagram}) has some implication concerning the description of the coupling between the different phases. The Landau free energy for the two order parameters can be written as \cite{bourdarot,chandra, chandra_2}
\begin{equation}
\label{op_coupling}
F=F_{\psi}+F_m+\gamma\psi m+g\psi^2m^2 ,
\end{equation}
where $\psi$ and $m$ denote the order parameters for the hidden order and the magnetic phase, respectively. A situation with a non-zero coupling term $\gamma\psi m$ would imply that both order parameters appear simultaneously when crossing a second order line from the paramagnetic state and that the first order line terminates at a critical point $p_c$, below the second order line (see \fref{uru2si2_schematic}). The present $\mu$SR data clearly indicate a phase separation and the merging of the first order line with the second order lines pointing therefore to a situation with $\gamma=0$ and suggesting that the order parameters $m$ and $\psi$ do not transform according to the same irreducible representation, underlining therefore their different nature. Such observation seems compatible with theories describing the hidden order state by an ordered phase of quadrupoles with order parameter $O_{x^2-y^2}$ (or $O_{xy}$) contrasting with the ordered dipoles of the magnetic phase with order parameter $O_z$ \cite{ohkawa}. As noted by Amitsuka \etal \cite{amitsuka_2}, in this scenario the relation
between $O_z$ and $O_{x^2-y^2}$ is equivalent to that between $S_z$ and $S_x$($S_y$) of the pseudo spin $S=\small{\frac{1}{2}}$. Therefore dipoles naturally occur when quadrupoles tilt, {\it i.e.} may appear as
domain boundaries of the hidden order phase and explain the isotropic field distribution observed by NMR and $\mu$SR at ambient pressure. Alternatively, the hidden order phase was also described in term of orbital magnetism \cite{chandra_2,chandra_3}, whose extended current loops could possibly produce an apparent isotropic field distribution at the silicon and muon site. 

In summary, the relation between antiferromagnetic order and hidden order in URu$_2$Si$_2$ has been discussed on the basis of recent $\mu$SR studies. From the observed behaviour under pressure, it is demonstrated that the two ordered states are not coupled but rather phase separated with almost degenerated condensing energies. Although the presented data show a body of evidence that the hidden order phase is connected to the observation of an isotropic field distribution, open questions remain concerning the true nature of the hidden order phase and the mechanism of its competition with the magnetic phase state. 

\section{Conclusions}
The studies, presented above, performed on the heavy-fermion superconductors exhibiting weak magnetic moments, are exemplary of the particular and complementary information which can be obtained by $\mu$SR. On the systems U$_{1-x}$Th$_x$Pt$_3$ and U(Pt$_{1-x}$Pd$_x$)$_3$, both the sensitivity to small moments and the different time window compared to neutron diffraction studies were utilized to demonstrate that static magnetism does not coexist with heavy-fermion superconductivity. In addition, the observation of critical points in the phase diagram of U(Pt$_{1-x}$Pd$_x$)$_3$ points to a superconducting triplet state and suggests the possible role played by ferromagnetic fluctuations for the heavy-fermion superconducting state in UPt$_3$.  Similar information was deduced for URu$_2$Si$_2$, where the ability of $\mu$SR to measure magnetic volume fraction was exploited to show that the magnetic phase is absent at ambient pressure and therefore does not coexist with heavy-fermion superconductivity. On this system, more fundamental comprehension of the peculiar relation between hidden order phase and magnetic phase could be gained by $\mu$SR measurements under pressure, where a clear phase separation could be demonstrated.

The authors are grateful to C.~Baines, P.~Estrela, F.~Gygax, Y.K.~Huang, J.A.~Mydosh, C.P.~Opeil, A.~Raselli, K.~Tenya, M.~Yokoyama, U.~Zimmermann for valuable discussions and/or their help during the experiments. This work is based on measurements performed at the Swiss Muon Source, Villigen, Switzerland. We also acknowledge support obtained through the European Science Foundation-FERLIN research program and the Petroleum Research Fund of the American Chemical Society.

\Figures
\end{document}